\documentclass[conference]{IEEEtran}
\IEEEoverridecommandlockouts
\usepackage{cite}
\usepackage{amsmath,amssymb,amsfonts}
\usepackage{algorithmic}
\usepackage{graphicx} 
\usepackage{textcomp}
\usepackage{xcolor}
\usepackage[caption=false]{subfig}
\def\BibTeX{{\rm B\kern-.05em{\sc i\kern-.025em b}\kern-.08em
    T\kern-.1667em\lower.7ex\hbox{E}\kern-.125emX}}
\begin{document}

\title{Deep Joint Source Channel Coding With Attention Modules Over MIMO Channels
\thanks {This work is supported by the Beijing Natural Science Foundation (L211012); the Natural Science Foundation of China (62122012, 62221001).}}

\author{\IEEEauthorblockN{Weiran Jiang,
		Wei Chen,
		Bo Ai}
	\IEEEauthorblockA{School of Electronic and Information Engineering, Beijing Jiaotong University, Beijing, China}

Corresponding author: Wei Chen\\
Email: \{22120071, weich, boai\}@bjtu.edu.cn
	}

\maketitle

\begin{abstract}
In this paper, we propose two deep joint source and channel coding (DJSCC) structures with attention modules for the multi-input multi-output (MIMO) channel, including a serial structure and a parallel structure. With singular value decomposition (SVD)-based precoding scheme, the MIMO channel can be decomposed into various sub-channels, and the feature outputs will experience sub-channels with different channel qualities. In the serial structure, one single network is used at both the transmitter and the receiver to jointly process data streams of all MIMO subchannels, while data steams of different MIMO subchannels are processed independently via multiple sub-networks in the parallel structure. The attention modules in both serial and parallel architectures enable the system to adapt to varying channel qualities and adjust the quantity of information outputs with the channel qualities. Experimental results demonstrate the proposed DJSCC structures have improved image transmission performance, and reveal the phenomenon via non-parameter entropy estimation that the learned DJSCC transceivers tend to transmit more information over better sub-channels.
\end{abstract}

\section{Introduction}

Shannon's separation theorem states that when the length of the codeword is infinite, it is optimal to design communication systems with a source encoder and a channel encoder that are separate from each other \cite{cover1999elements}. In accordance with the theorem, researchers have been optimizing communication systems in the manner of separate source channel coding (SSCC) for many years, leading to rapid advancement from the first to the fifth generations of mobile communication systems. 

In the practical scenario with limited code length, Shannon's separation theorem is no longer applicable, and joint source and channel coding (JSCC) has been demonstrated to outperform SSCC \cite{2}. Designing JSCC optimally is not an easy task. By utilizing deep learning (DL), it is possible to gain knowledge of JSCC design from data. For example, in \cite{djscc-p, burth2020joint, djscc-f, ntscc, protection}, DL-based JSCC is applied for image transmission. Previous work has shown When the channel quality is not as good as anticipated, the traditional communication performance deteriorates drastically, while DJSCC degrades gracefully\cite{djscc-semcom}. This has sparked a lot of enthusiasm and further research, including DJSCC with channel feedback\cite{djscc-f}, DJSCC with attention modules (ADJSCC)\cite{adjscc}, DJSCC with learnable quantization layers\cite{djscc-q}, and so on. Particularlly, ADJSCC allows the neural network to adapt to changes in the channel. It inputs the channel quality into the attention modules, which modify the weights of the output features. Studies have demonstrated that ADJSCC can attain close to ideal performance in varying signal-to-noise ratios (SNRs) \cite{adjscc}. Various DJSCC network designs have been proposed in the literature based on the attention mechanism \cite{djscc-q,djsccofdm}. Furthermore, DL has also attracted broad attentions to deal with in CSI feedback \cite{csifb,mcsifb,plug-and-play}, massive access \cite{ma1,ma2}, channel coding \cite{cc}. 

Most of the existing work on DJSCC considers the single-input single-output (SISO) channel. For open-loop multi-input multi-output (MIMO) systems, a DJSCC design with space-time coding is proposed in \cite{stjscc}. For close-loop MIMO systems, a Vision Transformer-based DJSCC design is proposed in \cite{Vit}, which utilizes multi-head self-attention mechanism to adapt to different MIMO sub-channels. However, Vision Transformer-based designs are challenging to implement in resource-limited devices due to their high computational and memory requirements \cite{spvit}. In addition, Vision Transformer-based methods would cause high latency\cite{Li_2023_ICCV}, and thus more efficient JSCC designs for MIMO are desired for low-latency applications, such as autonomous driving and telemedicine.

In this article, we present adaptive DJSCC structures for MIMO systems, which require a relatively small amount of parameters. We employ attention modules in the design. Specifically, we investigate both the serial and parallel network structures for MIMO. With singular value decomposition (SVD)-based precoding scheme, the MIMO channel can be decomposed into various sub-channels, and the feature outputs will experience sub-channels with different channel qualities. In the serial structure, one single network is used at both the transmitter and the receiver to jointly process data streams of all MIMO subchannels, while data steams of different MIMO subchannels are processed independently via multiple sub-networks in the parallel structure. We compare the two DJSCC structures in both the open-loop case \cite{stjscc} and the closed-loop case \cite{Vit}. Experiments demonstrate that both of the proposed DJSCC structures achieve improved performance. Furthermore, we reveal the phenomenon via non-parameter entropy estimation that the learned DJSCC transceivers tend to transmit more information over better sub-channels. 

The remainder of this paper is structured as follows. Section \ref{2} introduces the MIMO system model. The proposed serial and parallel DJSCC structures are presented in Section \ref{3}. Simulation results based on image transmission task are presented in Section \ref{4}. Finally, Section \ref{5} concludes this paper.

\section{System Model} \label{2}
\begin{figure}[t]
	\centerline{\includegraphics[width=0.45\textwidth]{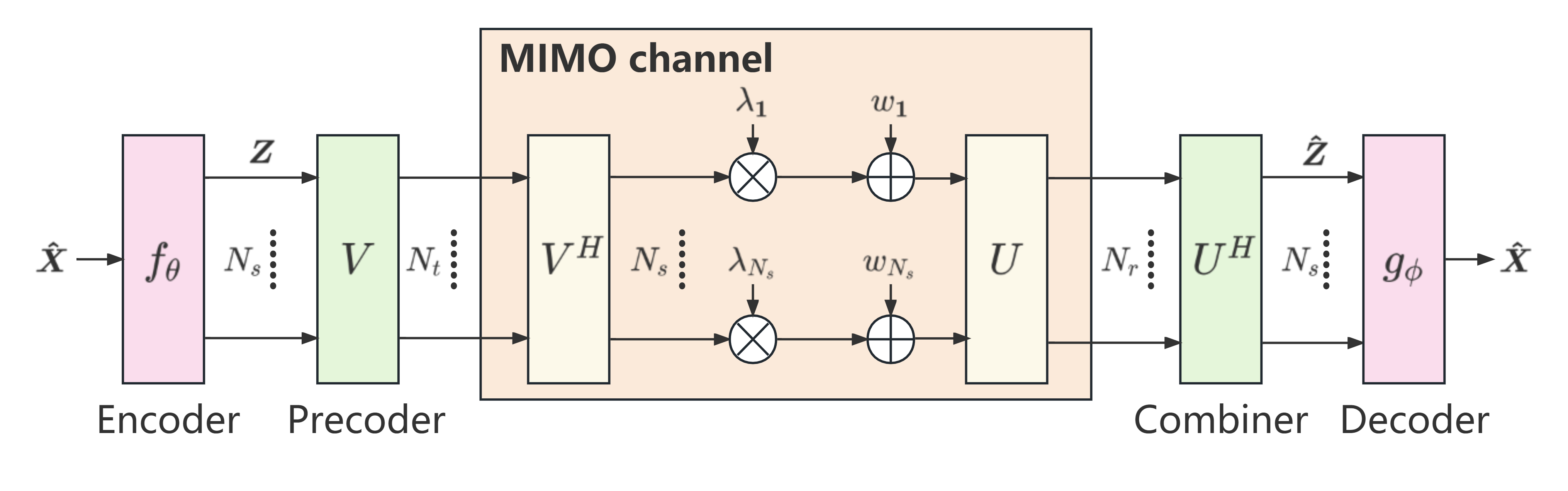}}
	\caption{The MIMO communication system with SVD-based precoding.}
	\label{fig1}
\end{figure}
We consider a MIMO system with precoding, where the channel state information (CSI) is known to both the transmitter and the receiver. The image signal is presented as a tensor $\boldsymbol{X}\in \mathbb{R} ^{C\times H\times W}$, where $H$ and $W$ are the height and width of the image, respectively, and $C$ represents the number of channels. As shown in Fig.~\ref{fig1}, we first map the source $\boldsymbol{X}$ to a low-dimensional feature matrix $\boldsymbol{Z}\in\mathbb{C} ^{N_s\times k}$ using a JSCC encoder, where $N_s$ denotes the number of data streams and $k$ is the number of channel uses. We define the $\emph{bandwidth ratio}$ as $\rho \triangleq k/CHW$. Before transmission, the feature matrix $\boldsymbol{Z}$ will be scaled to satisfy the power constraint $P$. Then MIMO precoding is applied to the features $\boldsymbol{Z}$ at the transmitter. At the receiver, the reverse operation is performed. We first recover $\boldsymbol{\hat{Z}}$ from the received signal and then reconstruct the image $\boldsymbol{\hat{X}}$ by a JSCC decoder. Our goal is to minimize the gap between the recovered image $\boldsymbol{\hat{X}}$ and the original image $\boldsymbol{X}$.

\subsection{MIMO channel and SVD-based precoding}
The MIMO channel can be represented as a matrix $\boldsymbol{H} \in \mathbb{C} ^{N_r \times N_t}$, where the transmitter has $N_t$ antennas and the receiver has $N_r$ antennas. We consider the block fading Rayleigh channel with sufficiently many scatters and no dominating scatter component. Thus, we can assume that elements of the channel matrix $\boldsymbol{H}$ follow a complex Gaussian distribution with zero mean and unit variance, and remains constant over the time of one feature matrix transmission. The process of transmitting feature matrix $\boldsymbol{Z}$ can be represented as 
\begin{equation}
	\boldsymbol{Y}=\boldsymbol{HZ}+\boldsymbol{W},
\end{equation}
where the additive white Gaussian noise $\boldsymbol{W} \in \mathbb{C}^{N_r \times k}$ follows $\boldsymbol{W}\sim \mathcal{CN}(\mathbf{0},\sigma^2\mathbf{I})$.

SVD-based precoding has been proven to achieve MIMO channel capacity in closed-loop case. In specific, the MIMO matrix $\boldsymbol{H}$ is decomposed via SVD as $\boldsymbol{H} = \boldsymbol{U \Sigma V}^H$, where $\boldsymbol{U} \in \mathbb{C}  ^{N_r\times N_r}$ and $\boldsymbol{V} \in \mathbb{C} ^{N_t\times N_t}$ are unitary matrices, and $\boldsymbol{\Sigma} \in \mathbb{C} ^{N_r\times N_t}$ is rectangular diagonal matrix with non-negative diagonal elements $\lambda_i$ in descending order. By applying the precoding matrix $\boldsymbol{V}$, the MIMO communication model can be represented as
\begin{equation}
	\boldsymbol{Y}=\boldsymbol{U \Sigma V}^H \boldsymbol{V Z}+\boldsymbol{W}.\label{MIMO_1}
\end{equation}

Multiplying the combiner matrix $\boldsymbol{U}^H$ with the received signal $\boldsymbol{Y}$, the equivalent MIMO channel becomes $\boldsymbol{\Sigma}$. By denoting $\boldsymbol{\hat{Z}}=\boldsymbol{U}^H\boldsymbol{Y}$ and $\boldsymbol{\hat{W}}=\boldsymbol{U}^H\boldsymbol{W}$, the MIMO model \eqref{MIMO_1} can be simplified into
\begin{equation}
	\boldsymbol{\hat{Z}}=\boldsymbol{\Sigma Z}+\boldsymbol{\hat{W}}. \label{MIMO_2}
\end{equation}
As $\boldsymbol{W}$ is a Gaussian random matrix and $\boldsymbol{U}^H$ is a complex orthogonal matrix, $\boldsymbol{\hat{W}}$ follows the same distribution as $\boldsymbol{W}$. As the equivalent channel matrix $\boldsymbol{\Sigma }$ is a rectangular diagonal matrix, the MIMO channel can be seen as $N_s$ independent sub-channels as shown in Fig.~\ref{fig1}, where $N_s$ ($N_s\leq \min\{N_r, N_t\}$) denotes the number of non-zero singular values in $\boldsymbol{\Sigma}$. Hence the $i$th row of feature matrix $\boldsymbol{Z}$ transmits over the $i$th sub-channel can be expressed as
\begin{equation}
	\boldsymbol{\hat{z}_i}=\lambda_i \boldsymbol{z_i}+\boldsymbol{\hat{w}_i}, \label{MIMO_3}
\end{equation}
where $\boldsymbol{z_i}\in\mathbb{C} ^{1\times k}$ and $\boldsymbol{\hat{z_i}}\in\mathbb{C} ^{1\times k}$ denote the transmitted and received feature vectors, respectively, $\boldsymbol{\hat{w}_i}\in\mathbb{C} ^{1\times k}$ denotes additive noise over the $i$th sub-channel, and $\lambda_i$ denotes the $i$th singular value corresponding to the $i$th sub-channel. Thus, the equivalent signal-to-noise ratio (SNR) of the $i$th sub-channel can be calculated as 
\begin{equation}
	SNR_i=\frac{\lambda_i^2 P}{N_s\sigma^2}. \label{SNR}
\end{equation}
Now the MIMO channel is decomposed into multiple SISO channels with different SNRs. The design of JSCC for MIMO system is more challenging, partly because features are transmitted via multiple sub-channels with different qualities.

\section{Proposed MIMO-DJSCC Structures with\\ Attention Modules } \label{3}
\begin{figure}[tbp]
	\centering
	\subfloat[The serial structure.]{\includegraphics[width=0.45\textwidth]{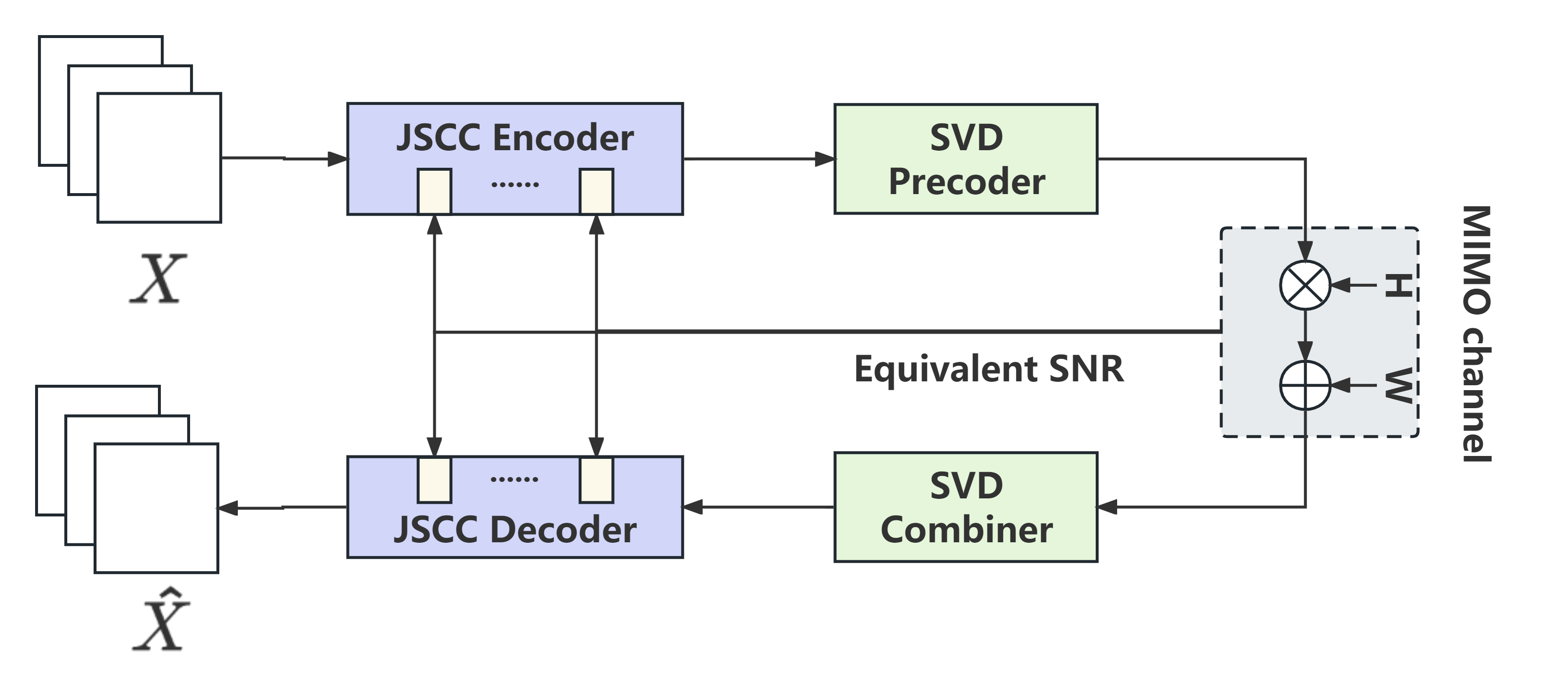}} 
	\\
	\subfloat[The parallel structure.]{\includegraphics[width=0.45\textwidth]{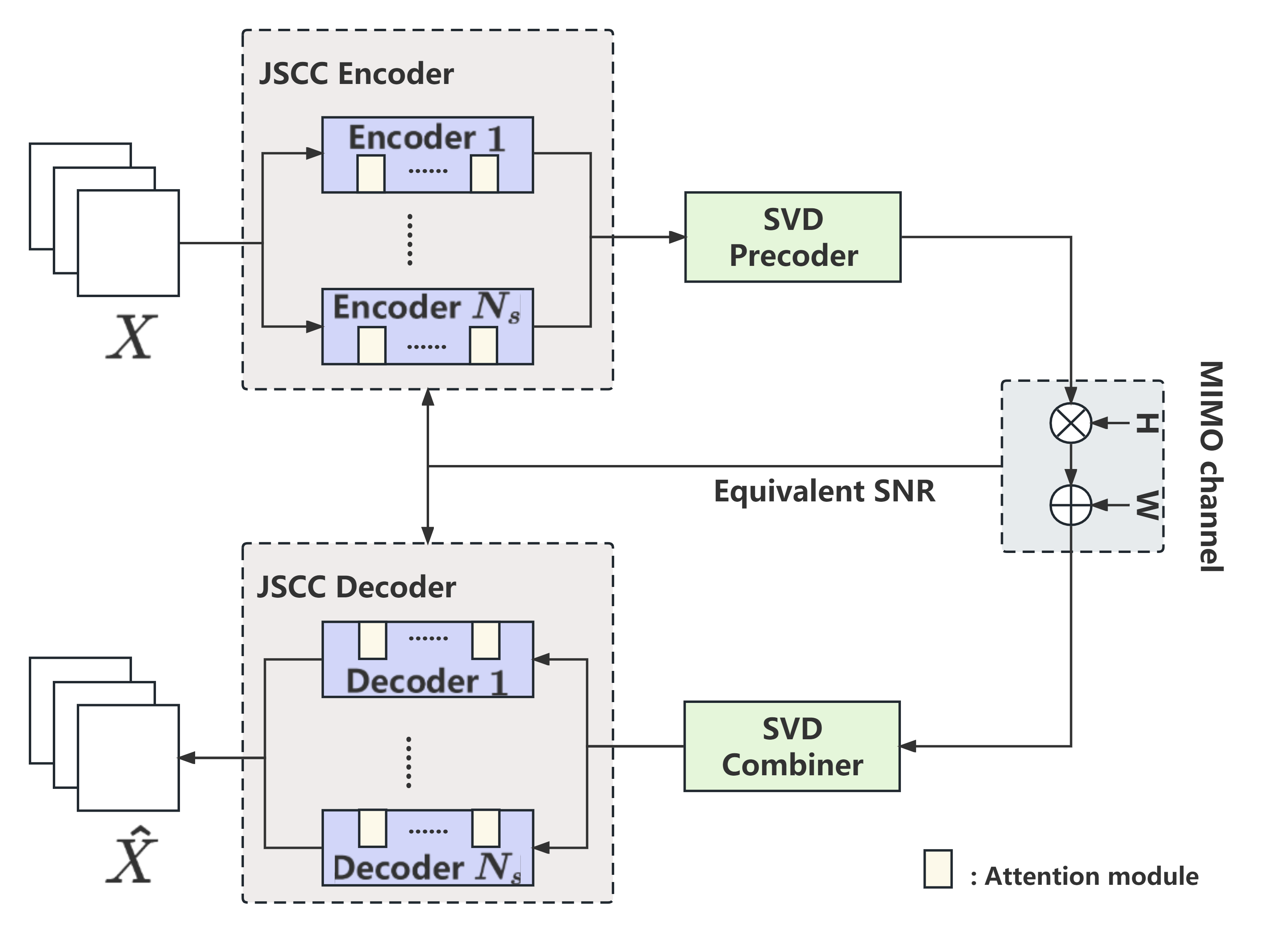}}
	\caption{The proposed MIMO-DJSCC structures.}
	\label{network}
\end{figure}

The ADJSCC has shown its capability to adjust to varying CSI in the SISO case \cite{adjscc}. As shown in Fig.~\ref{network}, we propose two MIMO-DJSCC network structures with attention modules, i.e., one serial structure and one parallel structure, for multiple sub-channels of different qualities. 

With perfect CSI, the equivalent SNRs of each sub-channels are given in \eqref{SNR}. In our proposed MIMO-DJSCC structures, we feed the equivalent SNRs into attention modules in both the encoder and the decoder. In the serial structure, the encoder network outputs features, which are then reshaped into an $N_s \times k$ matrix and multiplied by the precoding matrix. In contrast, the parallel structure contains multiple sub-networks of the same structure, and each sub-network is used for one MIMO subchannel. Then we concatenate the outputs of multiple sub-networks and multiply them with the precoding matrix. With the SVD-based precoder and combiner, the features from neural encoder will go through multiple independent sub-channels with different channel fading and noise. At the receiving end, the same serial/parallel structure is used to recover the original image. In this way, the serial structure outputs one single feature map of a high dimension, while the parallel structure with multiple sub-networks outputs multiple low dimensional feature maps.

\begin{figure}[t]
	\centerline{\includegraphics[width=0.5\textwidth]{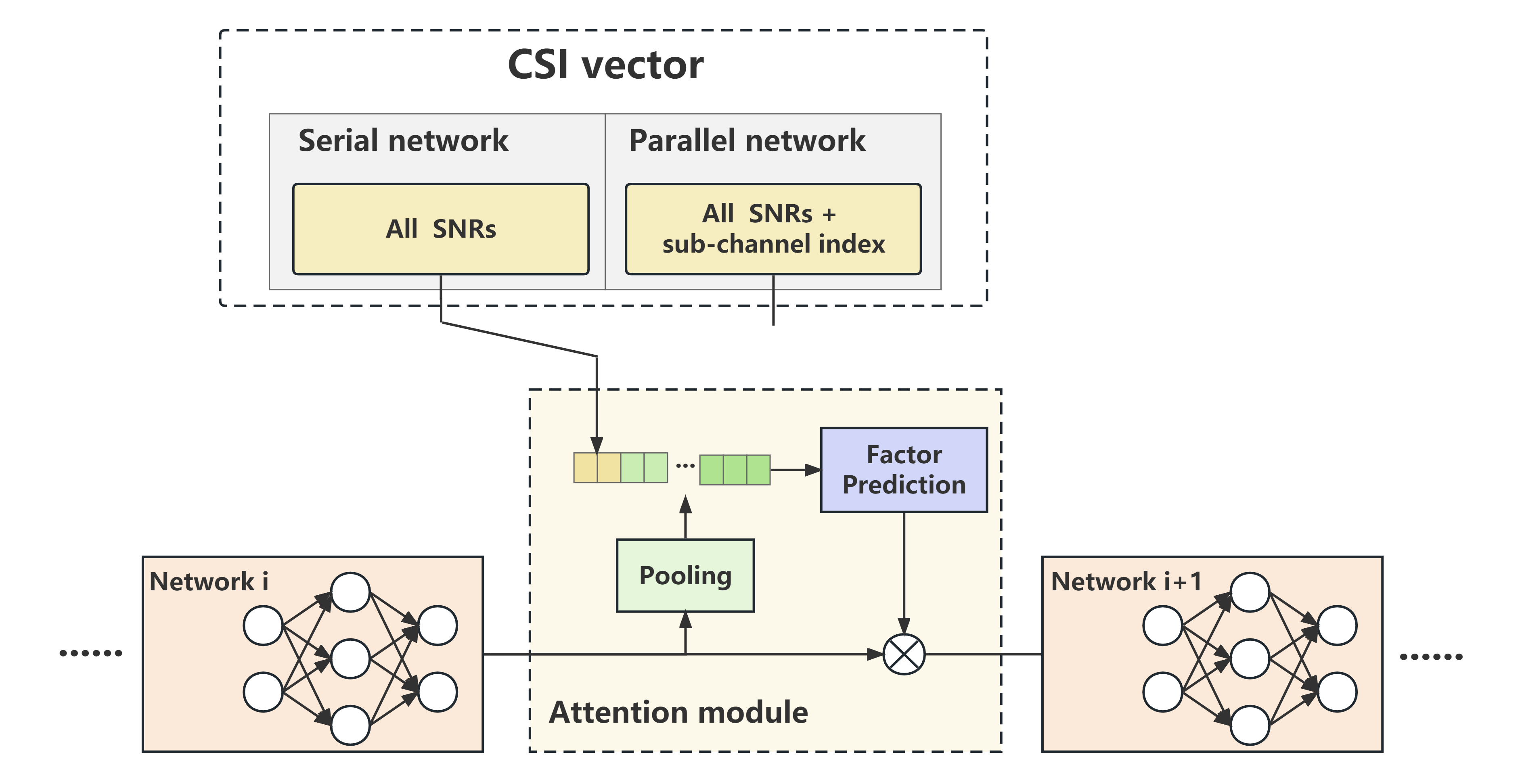}}
	\caption{The attention module in both the serial and parallel structures.}
	\label{attention}
\end{figure}

The attention module was first introduced into DJSCC in \cite{adjscc}, and then used for open-loop MIMO systems in\cite{stjscc}. For closed-loop MIMO, we use the attention modules to deal with multiple sub-channels of different qualities. The details of the attention module are shown in Fig.~\ref{attention}. In the serial structure, we input the equivalent SNRs of all sub-channels into the attention modules, since the feature vectors will go through all sub-channels. In the parallel structure, besides the equivalent SNRs, we input the sub-channel index vector into the attention module, represented by a one-hot vector for each sub-network. as the output features of each sub-network in the parallel structure only experience the corresponding sub-channel.

With the input CSI vector, attention modules learn to generate the weights for different features. First, a global average pooling layer pools the output features of the previous layer. The CSI vector is combined with the output vector from the pooling layer. The CSI vector contains the equivalent SNRs of all sub-channels and the corresponding index vectors if parallel structure is employed. We feed the extended vector into the factor prediction module, which consists of fully connected layers. This module utilizes both source features and channel-related information to generate a channel-wise soft attention vector for each source feature channel. We obtain the weight allocation by multiplying the attention vector with the source feature vector. Using the attention module is a cost-effective way to learn weight distribution, as only two linear layers with learnable parameters are required. The process can be expressed as follows:
\begin{equation}
\boldsymbol{F_{a}} =\boldsymbol{F_{s}} \cdot f_{\theta }(\text{concat} (\text{Pool} (\boldsymbol{F_{s}}),\boldsymbol{c})),
\end{equation}
where $\boldsymbol{F_{s}}$ denotes output features of the previous layer, $\boldsymbol{c}$ denotes the CSI vector, $\theta$ is the parameter set of the factor prediction layer, and $\boldsymbol{F_{a}}$ denotes the output feature.

In above, we provide detailed descriptions on the attention modules, as the capability of the DJSCC to adapt to different sub-channel quality comes from the attention modules, which take into account both the channel quality and the features of the source, and then assign different weights to the features. Unlike most previous models considered SISO channels, we propose two MIMO-DJSCC structures, i.e., serial and parallel structures, both of which employ feature learning layers and attention feature layers. In the experiments in Section \ref{4}, we simply use residual blocks in the feature learning module to showcase the performance, although other more complex feature learning modules can be applied in the proposed structure.

\section{Simulation Results} \label{4}
\begin{figure*}[tbp]
	\centering
	\begin{minipage}[t]{0.3\textwidth}
		\centering
		\includegraphics[width=\linewidth]{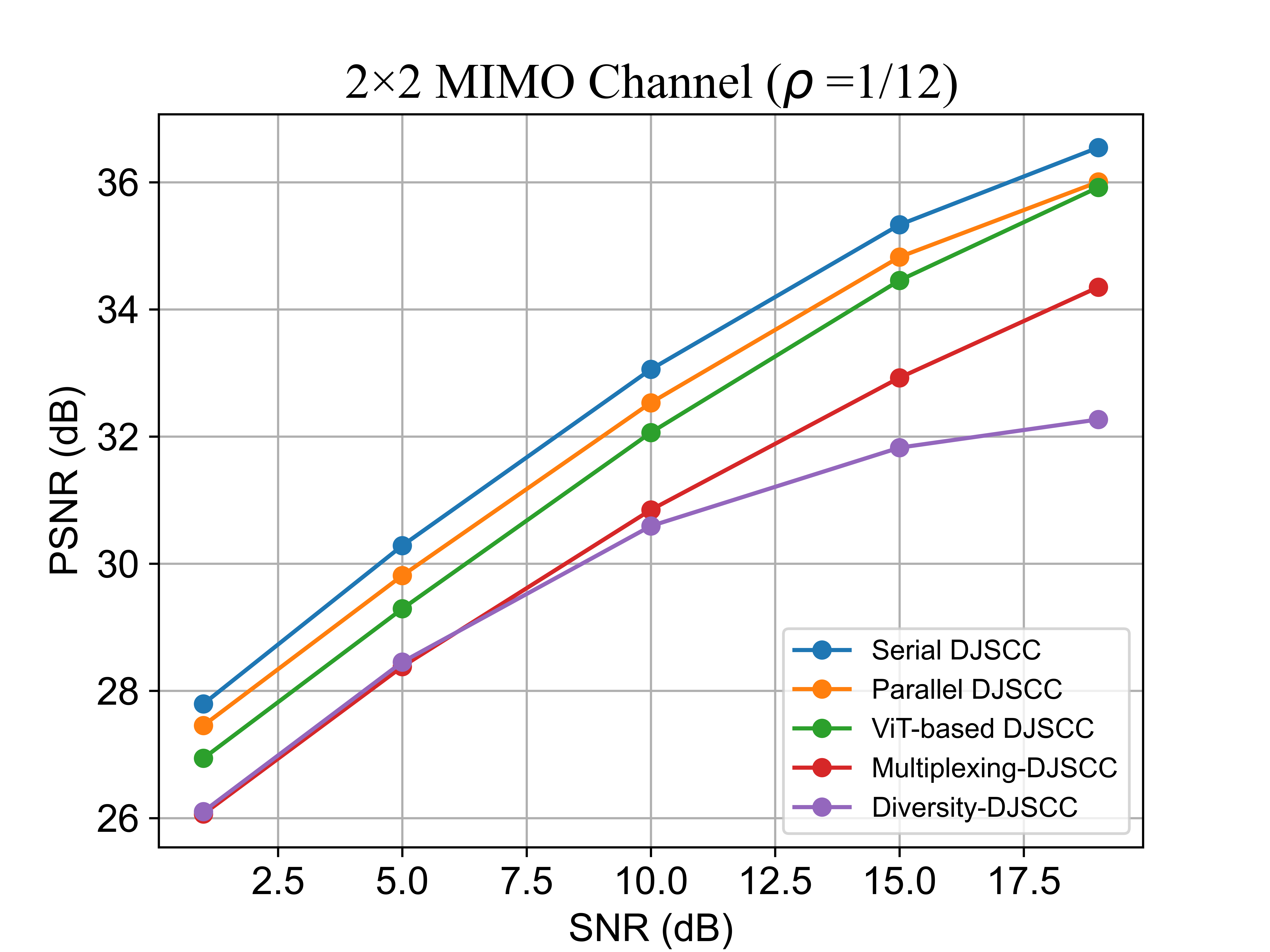}
		\caption{Performance of DJSCC schemes over MIMO channel. The SVD-based precoding is used in our serial and parallel MIMO-DJSCC and ViT-based DJSCC, and no precoding is used in Multiplexing/Diversity-DJSCC.}
		\label{psnr1}
	\end{minipage}
	\hspace{0.02\textwidth}
	\begin{minipage}[t]{0.3\textwidth}
		\centering
		\includegraphics[width=\linewidth]{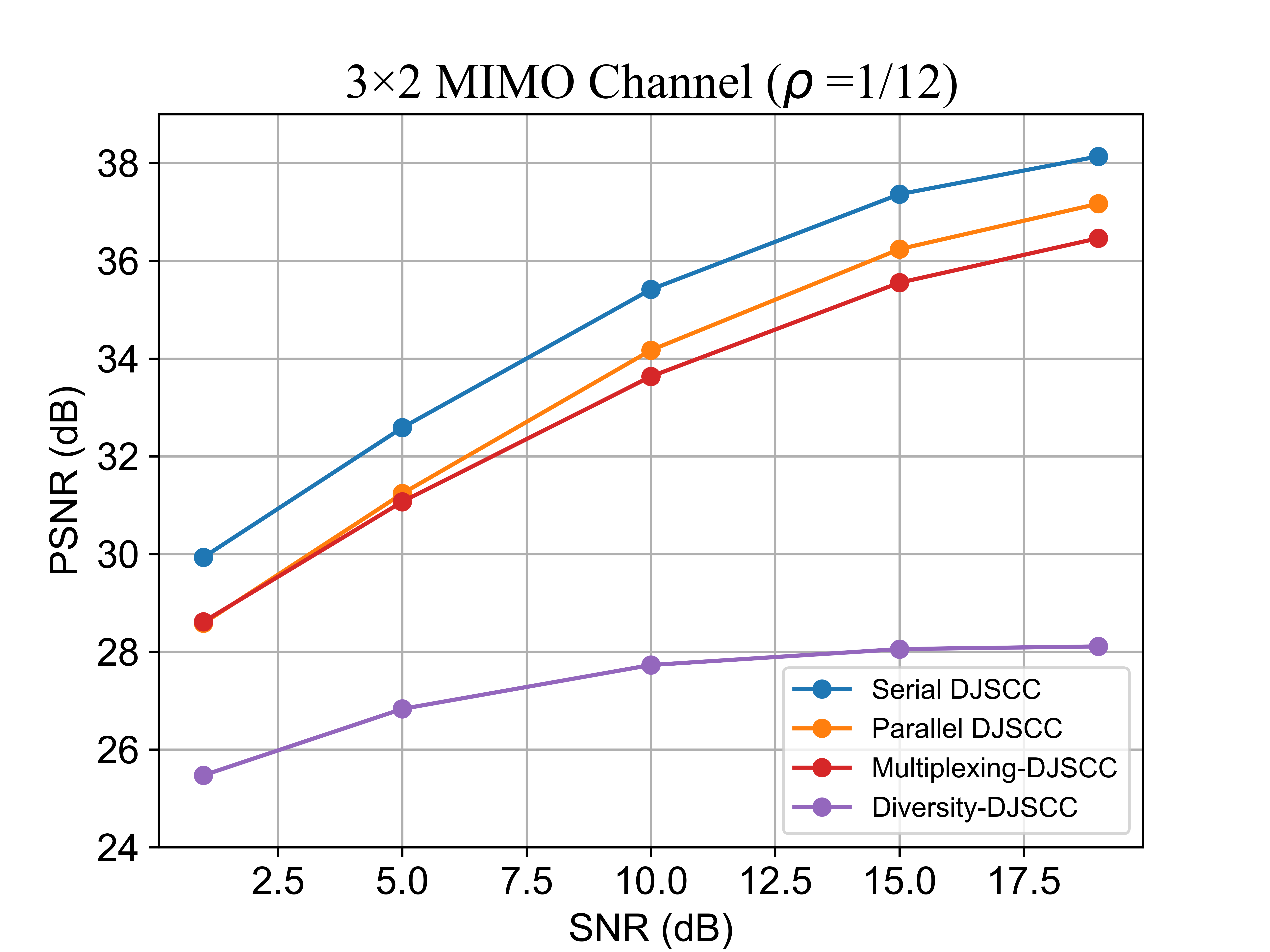}
		\caption{Performance of our serial and parallel MIMO-DJSCC and Multiplexing/Diversity-DJSCC over MIMO channel with $3$ transmitting antennas.}
		\label{psnr2}
	\end{minipage}
	\hspace{0.02\textwidth}
	\begin{minipage}[t]{0.3\textwidth}
		\centering
		\includegraphics[width=\linewidth]{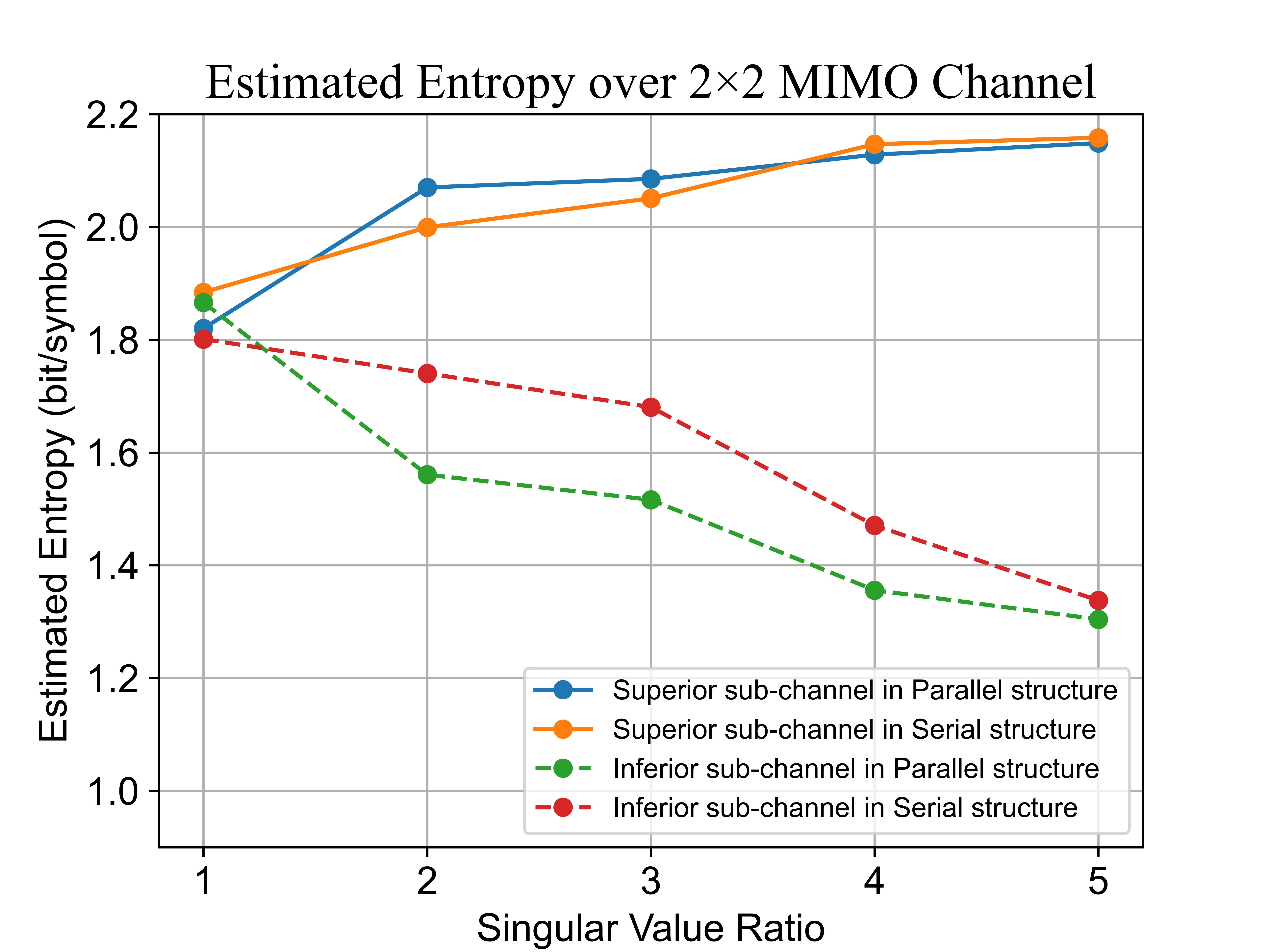}
		\caption{Estimated entropy of DJSCC encoder output over different sub-channels.}
		\label{entropy}
	\end{minipage}
\end{figure*}
In this section, we conduct experiments to evaluate the proposed MIMO-DJSCC methods. The numbers of transmitting antennas and receiving antennas are $N_t=2,3$, $N_r=2$, respectively. The channel matrix $\boldsymbol{H}$ is generated from complex Gaussian process, i.e. elements following $\boldsymbol{H}[i,j] \sim \mathcal{CN}(0,1) $. 

We use CIFAR-10 dataset in the evaluation. The batch size is set as 32, and the initial learning rate is set as $10^{-4}$ and gradually decreases with the progress of the training. The proposed networks are compared with both the scheme without precoding\cite{stjscc} and the scheme with precoding\cite{Vit} in MIMO scenario. For a fair comparison, we consider a network structure similar to that of \cite{stjscc}, which is composed of residual blocks as feature learning layers. Most computational cost comes from the feature learning layer compared to the attention feature layer. The loss function is defined as follows:
\begin{equation}
	\mathcal{L}=\frac{1}{N} \sum_{i}^{N} d(\boldsymbol{X}_i,\hat{\boldsymbol{X}}_i ), \label{loss}
\end{equation}
where $N$ denotes the number of image samples and $d(\boldsymbol{X}_i,\hat{\boldsymbol{X}}_i )=\frac{1}{n} \left \| \boldsymbol{X}_i-\hat{\boldsymbol{X}}_i  \right \|_F^2 $ denotes the mean squared error (MSE) between image sample $\boldsymbol{X}_i$ and the reconstructed image $\hat{\boldsymbol{X}}_i$, and $n$ denotes the number of elements in $\boldsymbol{X}_i$. 

\subsection{Image reconstruction performance}
In order to compare the performance of the schemes with and without precoding, we examine the quality of the recovered images at the receiver and use the peak signal-to-noise ratio (PSNR) as a measure. The PSNR is defined as:
\begin{equation}
	\text{PSNR} = 10\log_{10}{\frac{\text{MAX}^2}{\text{MSE}}}(dB), \label{psnr} 
\end{equation}
where $\text{MAX}$ denotes the maximum possible value of the image pixels. The MSE is defined as $\text{MSE}=d(\boldsymbol{X}_i,\hat{\boldsymbol{X}}_i)$.

We first compare the proposed MIMO-DJSCC schemes, i.e., the serial and parallel networks, with i) DJSCC methods without precoding, i.e., the diversity and multiplexing schemes proposed in \cite{stjscc} and ii) DJSCC method with precoding, i.e., the Vision Transformer (ViT)-based DJSCC scheme using SVD-based precoding \cite{Vit}. 
In \cite{stjscc}, the diversity scheme employs the Alamouti coding for space-time block code at the transmitter and receiver, while the multiplexing scheme uses MMSE with a residual connection layer for equalization at the receiver. The performance comparison is shown in Fig.~\ref{psnr1}. We generate the dataset with $SNR_{train}$ uniformly sampled in the range of $ \left [ 0,22 \right ] $ dB, and set the bandwidth ratio $\rho = 1/12$. We obtain the performance of the diversity and multiplexing schemes by using the code provided by the original authors of \cite{stjscc}. We draw the PSNR performance of ViT-based DJSCC scheme according to the results reported in \cite{Vit}, as the experimental setup is the same to \cite{Vit}, and the source code of ViT-based DJSCC is not yet publicly available.

For $2\times 2$ MIMO channels, it can be seen from Fig.~\ref{psnr1} that the DJSCC schemes with precoding, i.e., our proposed MIMO-DJSCC and ViT-based DJSCC \cite{Vit}, which requires CSI at the transmitter, have better image recovery performance than schemes without precoding at all SNRs. The results show the significant benefit from precoding. In comparison to the ViT-based DJSCC scheme with precoding, it is surprising that our schemes have better PSNR performance in this task, in view of the fact that the transformer has achieved the best performance in many applications. It would be interesting to explore the causes when the source code of ViT-based DJSCC is publicly available. Furthermore, the serial scheme performs better than the parallel scheme, which has fewer latent variables and can be seen as a special case of the serial scheme. 

We now increase the number of transmitting antennas to $3$, while keeping all other settings the same. The experimental results are shown in Fig.~\ref{psnr2}. The proposed MIMO-DJSCC schemes still have higher PSNRs compared to DJSCC without precoding. Compared to the results shown in Fig.~\ref{psnr1}, it is observed that the multiplexing scheme without precoding has improved PSNRs with increasing transmitter antennas, while the diversity scheme obtains negative gain with increasing transmitter antennas. The reason is given in \cite{stjscc} that as the number of transmitting antennas grows, the code rate of the Alamouti scheme declines, and thus the length of the feature vector length is reduced. However, with the increase of transmitter antennas, the serial and parallel networks achieve an average of 2.10 dB and 1.35 dB enhancement in PSNR, respectively. According to the comparisons, the proposed serial and parallel DJSCC with precoding can benefit from the increase of antennas in MIMO systems.

\subsection{Entropy of DJSCC encoder output over sub-channels}
The MIMO channel with SVD can be regarded as multiple sub-channels with different channel qualities and capacity. However, it is not clear, at least not in a quantified manner, how the sub-channel quality affects the amount of information transmitted in the DJSCC sub-channel.

The biggest issue is that there is no unified definition of semantic entropy, and the definitions that have been proposed are not as general as the Shannon entropy\cite{beyondbits}.
In order to investigate how much information is transmitted over different DJSCC sub-channels, we exploit the non-parameter entropy estimation method to estimated differential entropy of DJSCC encoder output, which related to the shortest description length\cite{cover1999elements}. Therefore, we use differential entropy to approximately quantify the amount of transmitted information.

The goal of non-parameter entropy estimation is to estimate the differential entropy of the underlying distribution only from a finite number of samples\cite{knn2016}. In our experiment, we adopt the kNN entropy estimation method, whose error rate is no more than twice of the Bayesian error rate (minimum error). The core idea of kNN methods is to estimate the probability mass of local field with $k$ nearest points around the sample\cite{knn2016}. The calculation formula of estimated differential entropy $\widehat {H} (\boldsymbol{X})$ is given by
\begin{equation}
	 \widehat {H}  (\boldsymbol{X})=\psi(N)-\psi(k)+  \log ( c_ {d} )+ \frac {d}{N} \sum _ {i=1}^ {N} \log {(\varepsilon (i)}),
\end{equation}
where $\psi$ denotes the digamma function, $N$ denotes the total number of samples, $k$ denotes the number of samples in the local field, $c_d$ is a constant term that denotes the unit volume of the local field, $d$ denotes the dimension of the data, and $\varepsilon (i)$ denotes the distance from the $i$th sample to the $k$th nearest neighbor point. Details of the formula and derivation process can be found in\cite{knn2016}. In our simulation, we set $d = 2$ and $k=\sqrt{N}$ without loss of generality.

In order to observe how differential entropy changes with different sub-channel qualities, we generate MIMO channel matrices with fixed singular values , which satisfy the power constraint $ \sum \lambda _{i}^2 = C $. We consider $2 \times 2$ MIMO channel and the SNR of 10 dB. The estimated entropy are shown in Fig.~\ref{entropy}. When the sub-channels' singular value ratio $r=1$, the two sub-channels have the same equivalent SNR as well as channel capacity, and the data streams going through the two sub-channels also have very close differential entropy values. We can observe that the differential entropy of the encoder output follows the pattern: the higher the channel capacity, the higher the differential entropy, and vice versa. By exploiting the entropy estimation, we reveal the mechanism of DJSCC in different sub-channels, i.e., more information is transmitted in the better sub-channel.

\section{Conclusion} \label{5}
In this paper, we design two DJSCC structures with attention modules for the MIMO channel, including a serial structure and a parallel structure. We apply the MIMO-DJSCC schemes for image transmission, and both network structures show improved image recovery performance. Furthermore, we investigate the estimated entropy of MIMO-DJSCC encoder over different sub-channels and observe that more information is transmitted in the better sub-channel. The combination of attention modules and SVD-based precoding enables the network to adapt to different sub-channels, generate data streams of different entropy. Thus, the MIMO-DJSCC could exploit the spatial gain of the MIMO channel, that results in improved performance than the SISO channel.

\bibliographystyle{IEEEtran}
\bibliography{ref}

\begin{thebibliography}{10}
\providecommand{\url}[1]{#1}
\csname url@samestyle\endcsname
\providecommand{\newblock}{\relax}
\providecommand{\bibinfo}[2]{#2}
\providecommand{\BIBentrySTDinterwordspacing}{\spaceskip=0pt\relax}
\providecommand{\BIBentryALTinterwordstretchfactor}{4}
\providecommand{\BIBentryALTinterwordspacing}{\spaceskip=\fontdimen2\font plus
\BIBentryALTinterwordstretchfactor\fontdimen3\font minus \fontdimen4\font\relax}
\providecommand{\BIBforeignlanguage}[2]{{%
\expandafter\ifx\csname l@#1\endcsname\relax
\typeout{** WARNING: IEEEtran.bst: No hyphenation pattern has been}%
\typeout{** loaded for the language `#1'. Using the pattern for}%
\typeout{** the default language instead.}%
\else
\language=\csname l@#1\endcsname
\fi
#2}}
\providecommand{\BIBdecl}{\relax}
\BIBdecl

\bibitem{cover1999elements}
T.~M. Cover, \emph{Elements of information theory}.\hskip 1em plus 0.5em minus 0.4em\relax John Wiley \& Sons, 1999.

\bibitem{2}
F.~Zhai, Y.~Eisenberg, and A.~K. Katsaggelos, ``Joint source-channel coding for video communications,'' \emph{Handbook of Image and Video Processing}, pp. 1065--1082, 2005.

\bibitem{djscc-p}
E.~Bourtsoulatze, D.~B. Kurka, and D.~Gündüz, ``Deep joint source-channel coding for wireless image transmission,'' in \emph{ICASSP 2019 - 2019 IEEE International Conference on Acoustics, Speech and Signal Processing (ICASSP)}, 2019, pp. 4774--4778.

\bibitem{burth2020joint}
D.~Burth~Kurka and D.~G{\"u}nd{\"u}z, ``Joint source-channel coding of images with (not very) deep learning,'' in \emph{International Zurich Seminar on Information and Communication (IZS 2020). Proceedings}.\hskip 1em plus 0.5em minus 0.4em\relax ETH Zurich, 2020, pp. 90--94.

\bibitem{djscc-f}
D.~B. Kurka and D.~Gündüz, ``Deep joint source-channel coding of images with feedback,'' in \emph{ICASSP 2020 - 2020 IEEE International Conference on Acoustics, Speech and Signal Processing (ICASSP)}, 2020, pp. 5235--5239.

\bibitem{ntscc}
J.~Dai, S.~Wang, K.~Tan, Z.~Si, X.~Qin, K.~Niu, and P.~Zhang, ``Nonlinear transform source-channel coding for semantic communications,'' \emph{IEEE Journal on Selected Areas in Communications}, vol.~40, no.~8, pp. 2300--2316, 2022.

\bibitem{protection}
J.~Xu, B.~Ai, W.~Chen, N.~Wang, and M.~Rodrigues, ``Deep joint source-channel coding for image transmission with visual protection,'' \emph{IEEE Transactions on Cognitive Communications and Networking}, vol.~9, no.~6, pp. 1399--1411, 2023.

\bibitem{djscc-semcom}
J.~Xu, T.-Y. Tung, B.~Ai, W.~Chen, Y.~Sun, and D.~Gündüz, ``Deep joint source-channel coding for semantic communications,'' \emph{IEEE Communications Magazine}, vol.~61, no.~11, pp. 42--48, 2023.

\bibitem{adjscc}
J.~Xu, B.~Ai, W.~Chen, A.~Yang, P.~Sun, and M.~Rodrigues, ``Wireless image transmission using deep source channel coding with attention modules,'' \emph{IEEE Transactions on Circuits and Systems for Video Technology}, vol.~32, no.~4, pp. 2315--2328, 2022.

\bibitem{djscc-q}
T.-Y. Tung, D.~B. Kurka, M.~Jankowski, and D.~Gündüz, ``Deepjscc-q: Constellation constrained deep joint source-channel coding,'' \emph{IEEE Journal on Selected Areas in Information Theory}, vol.~3, no.~4, pp. 720--731, 2022.

\bibitem{djsccofdm}
H.~Wu, Y.~Shao, K.~Mikolajczyk, and D.~Gündüz, ``Channel-adaptive wireless image transmission with ofdm,'' \emph{IEEE Wireless Communications Letters}, vol.~11, no.~11, pp. 2400--2404, 2022.

\bibitem{csifb}
J.~Xu, B.~Ai, N.~Wang, and W.~Chen, ``Deep joint source-channel coding for csi feedback: An end-to-end approach,'' \emph{IEEE Journal on Selected Areas in Communications}, vol.~41, no.~1, pp. 260--273, 2023.

\bibitem{mcsifb}
W.~Chen, W.~Wan, S.~Wang, P.~Sun, G.~Y. Li, and B.~Ai, ``Csi-pppnet: A one-sided one-for-all deep learning framework for massive mimo csi feedback,'' \emph{IEEE Transactions on Wireless Communications}, pp. 1--1, 2023.

\bibitem{plug-and-play}
W.~Wan, W.~Chen, S.~Wang, G.~Y. Li, and B.~Ai, ``Deep plug-and-play prior for multitask channel reconstruction in massive mimo systems,'' \emph{IEEE Transactions on Communications}, pp. 1--1, 2024.

\bibitem{ma1}
Y.~Bai, W.~Chen, B.~Ai, and P.~Popovski, ``Deep learning for asynchronous massive access with data frame length diversity,'' \emph{IEEE Transactions on Wireless Communications}, pp. 1--1, 2023.

\bibitem{ma2}
S.~Liang, W.~Chen, Z.~Sun, A.~Chen, and B.~Ai, ``Cluster-based massive access for massive mimo systems,'' \emph{China Communications}, vol.~21, no.~1, pp. 24--33, 2024.

\bibitem{cc}
Y.~Cheng, W.~Chen, L.~Li, and B.~Ai, ``Rate compatible ldpc neural decoding network: A multi-task learning approach,'' \emph{IEEE Transactions on Vehicular Technology}, pp. 1--5, 2023.

\bibitem{stjscc}
C.~Bian, Y.~Shao, H.~Wu, and D.~Gunduz, ``Space-time design for deep joint source channel coding of images over mimo channels,'' in \emph{2023 IEEE International Workshop on Signal Processing Advances in Wireless Communication (SPAWC)}, 2023.

\bibitem{Vit}
H.~Wu, Y.~Shao, C.~Bian, K.~Mikolajczyk, and D.~G{\"u}nd{\"u}z, ``Vision transformer for adaptive image transmission over mimo channels,'' in \emph{ICC 2023 - IEEE International Conference on Communications}, 2023, pp. 1--6.

\bibitem{spvit}
Z.~Kong, P.~Dong, X.~Ma, X.~Meng, W.~Niu, M.~Sun, X.~Shen, G.~Yuan, B.~Ren, H.~Tang \emph{et~al.}, ``Spvit: Enabling faster vision transformers via latency-aware soft token pruning,'' in \emph{European Conference on Computer Vision}.\hskip 1em plus 0.5em minus 0.4em\relax Springer, 2022, pp. 620--640.

\bibitem{Li_2023_ICCV}
Y.~Li, J.~Hu, Y.~Wen, G.~Evangelidis, K.~Salahi, Y.~Wang, S.~Tulyakov, and J.~Ren, ``Rethinking vision transformers for mobilenet size and speed,'' in \emph{Proceedings of the IEEE/CVF International Conference on Computer Vision (ICCV)}, October 2023, pp. 16\,889--16\,900.

\bibitem{beyondbits}
D.~Gündüz, Z.~Qin, I.~E. Aguerri, H.~S. Dhillon, Z.~Yang, A.~Yener, K.~K. Wong, and C.-B. Chae, ``Beyond transmitting bits: Context, semantics, and task-oriented communications,'' \emph{IEEE Journal on Selected Areas in Communications}, vol.~41, no.~1, pp. 5--41, 2023.

\bibitem{knn2016}
D.~Lombardi and S.~Pant, ``Nonparametric k-nearest-neighbor entropy estimator,'' \emph{Physical Review E}, vol.~93, no.~1, p. 013310, 2016.

\end{thebibliography}
\end{document}